\documentclass{article}
\usepackage{arxiv}
\usepackage[utf8]{inputenc}
\usepackage[T1]{fontenc}
\usepackage{hyperref}
\usepackage{url}
\usepackage{booktabs}
\usepackage{amsfonts}
\usepackage{nicefrac}
\usepackage{microtype}
\usepackage{graphicx}
\usepackage{natbib}
\usepackage{doi}
\usepackage{caption}

\title{Radar DataTree: A FAIR and Cloud-Native Framework for Scalable Weather Radar Archives}

\author{
  \href{https://orcid.org/0000-0001-8081-7827}{\includegraphics[scale=0.06]{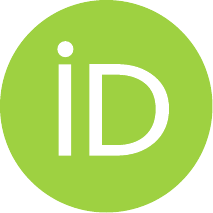}\hspace{1mm}Alfonso Ladino-Rincon} \\
  University of Illinois Urbana-Champaign \\
  \texttt{alfonso8@illinois.edu} \\
  \And
  \href{https://orcid.org/0000-0003-0348-0452}{\includegraphics[scale=0.06]{figures/orcid.pdf}\hspace{1mm}Stephen W. Nesbitt} \\
  University of Illinois Urbana-Champaign \\
  \texttt{snesbitt@illinois.edu} \\
}

\renewcommand{\shorttitle}{Radar DataTree}

\hypersetup{
pdftitle={Radar DataTree: A FAIR and Cloud-Native Framework for Scalable Weather Radar Archives},
pdfauthor={Alfonso Ladino, Stephen W. Nesbitt},
}

\begin{document}
\maketitle

\begin{abstract}
We introduce \textbf{Radar DataTree}, the first dataset-level framework that extends the WMO FM-301 standard from individual radar volume scans to time-resolved, analysis-ready archives. Weather radar data are among the most scientifically valuable yet structurally underutilized Earth observation datasets. Despite widespread public availability, radar archives remain fragmented, vendor-specific, and poorly aligned with FAIR (Findable, Accessible, Interoperable, Reusable) principles—hindering large-scale research, reproducibility, and cloud-native computation. Radar DataTree addresses these limitations with a scalable, open-source architecture that transforms operational radar archives into FAIR-compliant, cloud-optimized datasets. Built on the FM-301/CfRadial 2.1 standard and implemented using \texttt{xarray.DataTree}, Radar DataTree organizes radar volume scans as hierarchical, metadata-rich structures and serializes them to Zarr for scalable analysis. Coupled with \texttt{Icechunk} for ACID-compliant storage and versioning, this architecture enables efficient, parallel computation across thousands of radar scans with minimal preprocessing. We demonstrate significant performance gains in case studies including Quasi-Vertical Profile (QVP) and precipitation accumulation workflows, and release all tools and datasets openly via the \texttt{Raw2Zarr} repository. This work contributes a reproducible and extensible foundation for radar data stewardship, high-performance geoscience, and AI-ready weather infrastructure.
\end{abstract}

\keywords{FAIR data \and Weather radar \and Cloud-native \and ARCO \and Zarr \and Data stewardship}
\section{Introduction}
Ground-based weather radars are a cornerstone of atmospheric observation, capturing high-resolution data on precipitation, wind, and storm structures at fine temporal intervals. These systems support a wide range of societal and scientific applications, including severe weather forecasting, aviation safety, flood risk management, and climate-scale analysis. As radar networks expand globally and archives grow in volume and value, the potential for long-term, high-resolution atmospheric datasets continues to increase.

Yet despite their importance, radar data remains structurally challenging to use at scale. Existing archives are organized as millions of standalone binary files—often in proprietary or semi-standardized formats—designed for real-time operations, not scientific reuse. Metadata is inconsistently encoded, temporal indexing is absent, and no widely adopted framework currently supports organizing radar data as structured, versioned, and queryable datasets. Even open-access archives, such as NEXRAD Level II or Colombia's IDEAM radar network, present significant technical barriers to FAIR data use \citep{saltikoff2019overview}.

In this paper, we introduce the \textbf{Radar DataTree} framework: a cloud-native, FAIR-aligned data model and storage architecture that transforms fragmented radar files into scalable, structured datasets. Radar DataTree extends existing file-level standards (such as WMO FM-301 / CfRadial 2.1) to support time-aware, hierarchical organization of radar scans using the \texttt{xarray.DataTree} data model. The resulting datasets preserve the full sweep-level structure of each radar scan, while aligning collections along a shared time dimension for analysis.

Radar DataTree is implemented and deployed using the open-source \texttt{Raw2Zarr} pipeline \citep{ladino2025raw2zarr}. This end-to-end Python framework ingests raw radar files, decodes and restructures them using \texttt{Xradar}, and serializes the result to Zarr format using \texttt{Icechunk}—a transactional storage engine that provides ACID-compliant versioning and concurrent-safe cloud updates. This architecture enables scalable, reproducible workflows for multi-scan radar analysis, real-time ingestion, and machine learning pipelines.

We demonstrate the Radar DataTree framework on operational radar archives, including NEXRAD and SIGMET datasets, and benchmark its performance on core radar analysis tasks such as Quasi-Vertical Profile (QVP) generation and Quantitative Precipitation Estimation (QPE). Results show up to two orders of magnitude speedup over traditional file-based workflows with full support for cloud-native execution and reproducibility.

\textbf{This work contributes:}
\begin{itemize}
  \item A dataset-level data model for weather radar collections, compatible with FM-301 and Climate and Forecast (CF) conventions.
  \item An open-source implementation—\texttt{Raw2Zarr}—for ETL from operational radar archives into scalable, Icechunk-managed Zarr datasets.
  \item Empirical validation of the model’s performance and usability on real-world archives and scientific workflows.
\end{itemize}

By bridging radar-specific standards with modern data infrastructure, Radar DataTree provides a foundation for interoperable, scalable, and AI-ready weather radar science.
\section{Background and Motivation}

Weather radar systems produce some of the most spatiotemporally detailed observations in atmospheric science. Each radar volume scan captures reflectivity, Doppler velocity, and polarimetric variables across a three-dimensional cone of the atmosphere by rotating 360° at multiple fixed elevation angles. These scans are grouped into Volume Coverage Patterns (VCPs), which define the radar's sweep strategy—number of angles, rotation speed, and dwell time—based on the operational mode (e.g., clear-air or storm surveillance). Scans are typically repeated every 4–10 minutes, resulting in dense four-dimensional datasets \((x, y, z, t)\) that are critical for both real-time operations and retrospective analysis.

Despite their observational richness, radar archives are difficult to use due to how data is structured and stored. Most operational radar networks—including NEXRAD (U.S.), IDEAM (Colombia), and FMI (Finland)—store each VCP as an individual binary file in vendor-specific formats. These formats were designed for near-real-time visualization and have limited support for long-term reuse, metadata discovery, or spatial-temporal subsetting.

FM-301–compliant archives, while structurally improved, still require scan-by-scan parsing in the absence of a dataset-level abstraction. This file-centric model is fundamentally misaligned with FAIR data principles \citep{saltikoff2019overview}. While many archives are technically findable and accessible, they are rarely interoperable or reusable without expert knowledge and repeated effort.

Efforts like FM-301, Zarr, and \texttt{Xradar} have made strides in improving radar data structure. Radar DataTree builds upon this ecosystem to address what remains unsolved: a cohesive, time-aware, analysis-ready model for full radar collections.

\section{Related Work}

Radar data interoperability and standardization have long been active areas of development across both meteorological agencies and the open-source software community. Prior efforts have primarily focused on improving the usability of individual radar scans, with less emphasis on collection-level structure, temporal alignment, or cloud-native scalability.

\paragraph{File-level standards.} The most widely adopted specification for radar data is the WMO FM-301 standard, published as part of the Manual on Codes (WMO-No. 306) \citep{wmo2025fm301}. FM-301 defines a hierarchical schema for radar data based on the CfRadial~2.1 convention and CF metadata conventions. It allows individual radar sweeps to be encoded as NetCDF4 or HDF5 files with standardized variable names, coordinate systems, and metadata groups. However, FM-301 applies only at the level of single radar volume scans and does not specify how multiple files should be organized, indexed, or queried as a cohesive dataset over time.

\paragraph{Open-source radar libraries.} Several community-maintained Python packages provide tools for ingesting, decoding, and analyzing radar data. \texttt{Py-ART} offers a general-purpose radar processing toolkit that supports multiple formats and provides functionality for gridding, visualization, and filtering \citep{helmus2016pyart}. \texttt{Xradar} extends this ecosystem by providing strict support for FM-301/CfRadial~2.1 decoding and seamless integration with \texttt{xarray} objects \citep{xradar2025}. These libraries focus primarily on volume-level workflows and do not address scalable data storage, temporal indexing, or dataset-level modeling.

\paragraph{Cloud-native formats.} The geoscience community has increasingly adopted formats like Zarr for cloud-optimized, chunked storage of multidimensional arrays. Zarr's compatibility with object storage and parallel I/O has made it a preferred backend in frameworks such as Pangeo. However, raw Zarr stores lack native support for versioning, transactions, or structured group hierarchies unless coupled with additional tooling. Icechunk \citep{icechunk2024} addresses this gap by providing ACID-compliant transactional storage for Zarr datasets, enabling concurrent writes, atomic updates, and version-controlled metadata in distributed environments.

\paragraph{Community radar infrastructure.} The broader radar science community has increasingly embraced open-source software as a foundation for reproducible research and long-term data stewardship. Heistermann et al. \citep{heistermann2015openradar} surveyed the emergence of open-source radar tools, highlighting the role of community-led initiatives in lowering technical barriers and fostering interoperability.

\paragraph{Gap in dataset-level radar modeling.} To date, no framework combines radar-specific standards (e.g., FM-301) with scalable, transactional, time-aware modeling of entire radar archives. Existing tools operate primarily at the level of single files or in-memory arrays. Radar DataTree addresses this gap by providing a formal data model that represents radar collections as hierarchical, metadata-rich datasets aligned along time. By building on FM-301, implementing with \texttt{xarray.DataTree}, and persisting with Icechunk-managed Zarr stores, Radar DataTree enables scalable, reproducible radar workflows aligned with FAIR and cloud-native design principles.

To date, no framework combines radar-specific standards (e.g., FM-301) with scalable, transactional, time-aware modeling of entire radar archives. Existing tools operate primarily at the level of single files or in-memory arrays. Radar DataTree builds upon these efforts by providing a formal data model that represents radar collections as hierarchical, metadata-rich datasets aligned along time.

\section{Data Model and Architecture}

Radar data is inherently hierarchical and time-resolved. Each radar volume scan (VCP) consists of multiple sweeps at different elevation angles, with each sweep containing arrays of polarimetric variables (e.g., reflectivity, velocity, differential phase) on a polar grid. These scans repeat every few minutes, producing a nested, multivariate time series that spans days to decades. However, traditional radar archives store each scan as an isolated binary file, with no higher-level abstraction for organizing, aligning, or querying data over time.

\paragraph{Radar DataTree structure.} The \textbf{Radar DataTree} model addresses this limitation by representing a radar archive as a hierarchical, metadata-rich tree of datasets aligned along a common time axis. At its core, each radar volume scan is represented as a subtree containing its original FM-301 / CfRadial~2.1 group structure—preserving sweeps, coordinate metadata, and variable attributes. These subtrees are combined into a parent tree organized by observation time, forming a unified dataset that maintains both sweep-level detail and time-series navigability.

This structure is implemented using \texttt{xarray.DataTree}, a new hierarchical container in the xarray ecosystem that enables nested datasets with shared coordinate alignment and efficient traversal. Radar DataTree leverages this abstraction to organize heterogeneous scans into a coherent time-indexed archive while preserving full metadata integrity.

\paragraph{Metadata alignment and FM-301 compliance.} Each volume scan in the Radar DataTree preserves its original FM-301-compliant structure, including:

\begin{itemize}
  \item Nested groups for each sweep, encoded with consistent dimension names (e.g., time, range, azimuth)
  \item CF-compliant coordinate variables for spatial referencing
  \item Global and per-variable metadata for instrument settings, calibration, and units
\end{itemize}

To enable analysis across time, the Radar DataTree model introduces a top-level `time` dimension indexing all volume scans. This allows downstream tools to treat the archive as a temporally continuous dataset, with sweep-level metadata preserved per scan. 

\begin{table}[h]
\centering
\caption{Core components of the Radar DataTree ecosystem}
\begin{tabular}{ll}
\toprule
\textbf{Component} & \textbf{Role in Architecture} \\
\midrule
FM-301 & File-level standard for radar volumes and sweeps \\
\texttt{xarray.DataTree} & Hierarchical in-memory representation of scans \\
Zarr & Chunked, compressed, cloud-native storage format \\
Icechunk & ACID-compliant transactional engine for Zarr datasets \\
\bottomrule
\end{tabular}
\label{tab:ecosystem}
\end{table}

\paragraph{Zarr serialization and Icechunk persistence.} To support cloud-native access and scalable analytics, the Radar DataTree is serialized to the Zarr format. Zarr stores multidimensional arrays as compressed, chunked files in object storage, enabling efficient partial reads, lazy evaluation, and parallel processing. Each scan is stored as a group within a time-indexed hierarchy, preserving the internal sweep structure and associated metadata.

Radar DataTree archives are persisted using \texttt{Icechunk} \citep{icechunk2024}, a transactional storage engine that wraps Zarr stores with ACID guarantees. Icechunk maintains metadata catalogs, write-ahead logs, and atomic commit protocols to support safe concurrent access and version-controlled updates in distributed environments. This ensures that ingest pipelines, real-time workflows, and downstream users can operate on the same archive without risking data corruption or race conditions.

\begin{figure}[h]
  \centering
  \includegraphics[width=0.7\textwidth]{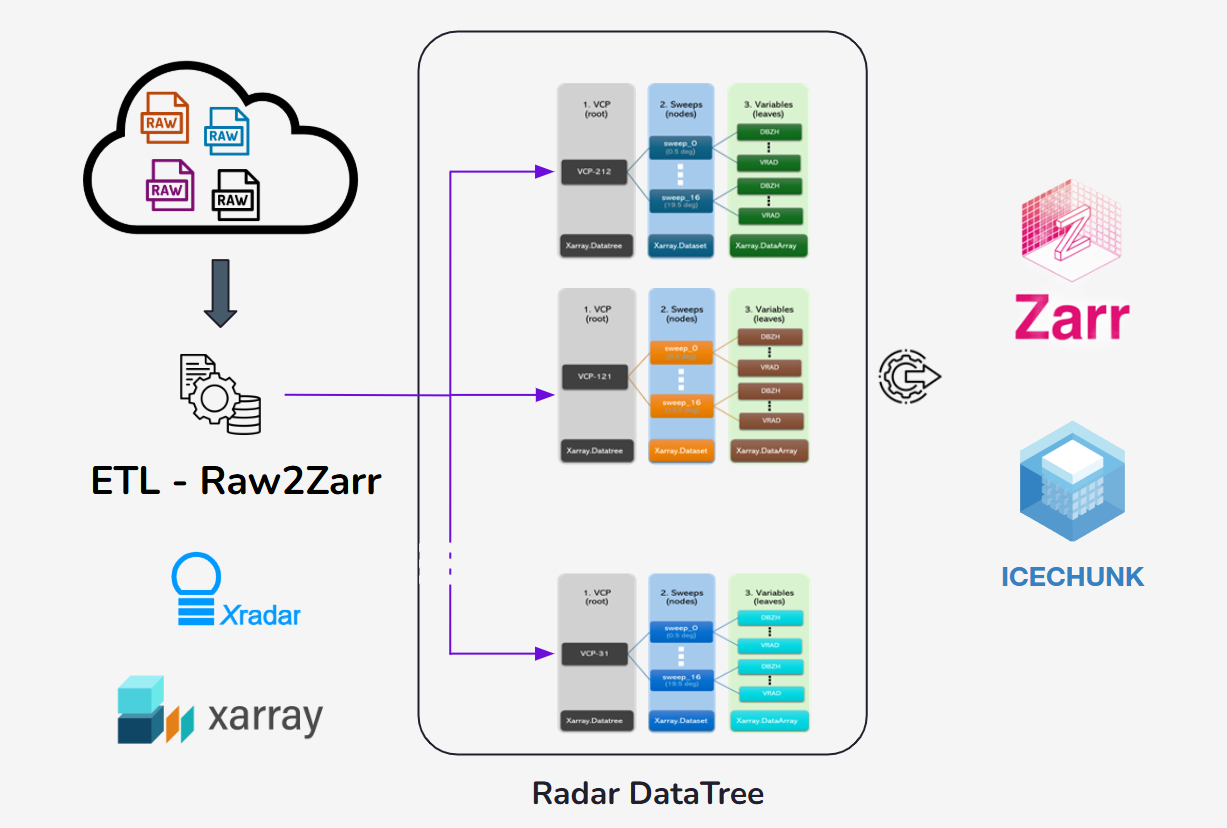}
  \caption{
    End-to-end ETL pipeline for radar archive ingestion. Raw radar files hosted in cloud storage (e.g., NEXRAD Level II, SIGMET) are decoded using \texttt{Xradar}, structured into a time-aligned hierarchical dataset using \texttt{xarray.DataTree}, and serialized into transactional \texttt{Zarr} stores managed by \texttt{Icechunk}. The entire process is implemented in the open-source \texttt{Raw2Zarr} package.
  }
  \label{fig:etl-datatree}
\end{figure}

\paragraph{Raw2Zarr pipeline.} The full architecture is operationalized in the open-source \texttt{Raw2Zarr} Python package \citep{ladino2025raw2zarr}, which performs end-to-end extraction, transformation, and loading (ETL) of radar archives, as illustrated in Figure~\ref{fig:etl-datatree}.

The pipeline consists of four stages:

\begin{enumerate}
  \item \textbf{Extraction:} Raw Level II or SIGMET files are downloaded from cloud buckets or S3-compatible object stores.
  \item \textbf{Transformation:} Files are decoded using \texttt{Xradar} \citep{xradar2025}, parsed into FM-301-compliant \texttt{xarray.Dataset} objects, and grouped by time.
  \item \textbf{Tree construction:} Each dataset is inserted as a node into a \texttt{DataTree}, preserving sweep metadata and scan structure.
  \item \textbf{Loading:} The complete tree is serialized to a Zarr store and committed to Icechunk for ACID-compliant persistence and version tracking.
\end{enumerate}

This design enables scalable, reproducible, and cloud-ready radar archives that can support a wide range of scientific and operational applications without repeated preprocessing or bespoke file handling.

\subsection{Interactive Access and Exploration in Practice}

\begin{figure}[h]
  \centering
  \includegraphics[width=0.95\textwidth]{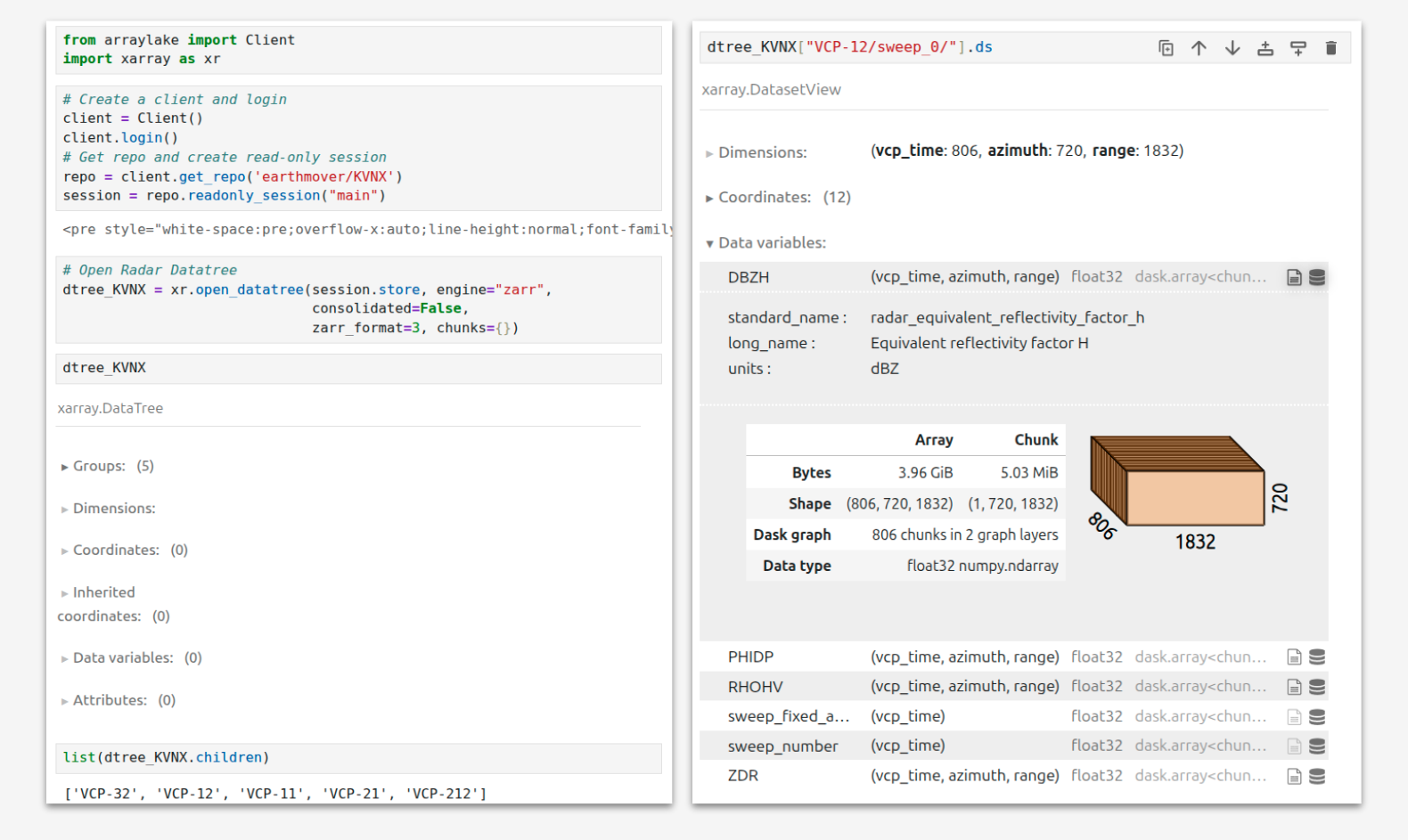}
  \caption{
  Interactive exploration of the KVNX Radar DataTree using \texttt{xarray.DataTree} and the \texttt{arraylake} Python client. The full May 2011 archive (\textasciitilde765~GB) is loaded as a single navigable object. Each node represents a Volume Coverage Pattern (VCP) and can be accessed using a simple path syntax. Radar reflectivity variables (e.g., DBZH) are self-described, chunked, and cloud-optimized for scalable analysis.
  }
  \label{fig:jupyter-kvnx}
\end{figure}

To demonstrate the practical use of the Radar DataTree model, Figure~\ref{fig:jupyter-kvnx} shows an interactive session accessing the full May 2011 archive of the KVNX radar site using the \texttt{xarray.DataTree} interface and the \texttt{arraylake} client. This archive spans over 765~GB of Zarr-encoded data persisted in an \texttt{Icechunk} store.

Once loaded, the radar collection is exposed as a time-indexed tree of datasets, with each node representing a distinct Volume Coverage Pattern (VCP). As shown in the interface, users can access any individual sweep using intuitive, path-like syntax (e.g., \texttt{"VCP-212/sweep\_0"}). Variables such as radar reflectivity (DBZH) are immediately accessible as self-described, chunked arrays with rich metadata, standard-compliant attributes, and an optimized layout for Dask-based processing.

This capability transforms radar data from a disjoint collection of binary files into a fully navigable scientific dataset. It enables:

\begin{itemize}
  \item Seamless subsetting and on-demand querying across large temporal ranges,
  \item Efficient use of cloud-native analysis tools and parallel processing,
  \item Reproducible exploration from within standard scientific Python environments such as Jupyter.
\end{itemize}

\section{Case Studies and Performance}

To evaluate the usability and efficiency of the Radar DataTree framework, we apply it to several common radar analysis workflows. These case studies highlight the benefits of dataset-level organization, parallel access, and cloud-native storage over traditional file-based methods.

\subsection{Quasi-Vertical Profile (QVP) Generation}

Quasi-Vertical Profiles (QVPs) summarize vertical trends in radar variables by compositing azimuthal means from constant-elevation sweeps over time \citep{ryzhkov2016qvp}. This technique provides insights into storm microphysics, melting layer detection, and hydrometeor classification.

In a traditional archive, each radar volume must be decoded independently, relevant sweep angles must be located and matched across files, and metadata inconsistencies handled manually. With Radar DataTree, QVP workflows are simplified: sweeps are uniformly stored, aligned in time, and queryable via a single interface.

On a one-week NEXRAD archive, QVP generation using Radar DataTree yielded over a 100× speedup compared to baseline Py-ART-based scripts operating on raw Level II files. Parallelized extraction using Dask enabled scalable compute across cloud-backed Zarr stores with minimal I/O overhead.

This demonstrates not only computational gains but also a substantial reduction in workflow complexity and code maintenance.

\subsection{Time-Series Extraction at Fixed Location}

Extracting radar time series at fixed spatial locations is essential for sensor intercomparisons, validation, and data assimilation workflows. Traditionally, this involves directory scanning, decoding binary files, coordinate transformation, and interpolation.

Radar DataTree abstracts these steps by treating each scan as a time-aligned, metadata-consistent node. Queries can be executed directly across the archive to extract pointwise time series without loading entire volumes into memory.

On a month-long NEXRAD archive, Radar DataTree reduced extraction latency and memory footprint by over an order of magnitude, allowing sub-minute retrieval of multi-week time series using cloud-native compute.

\subsection{Quantitative Precipitation Estimation (QPE)}

QPE uses reflectivity-derived rain rates to compute precipitation accumulations over time. One standard method applies the Marshall–Palmer Z–R relationship, which links reflectivity \( Z \) to rain rate \( R \) via an exponential distribution of drop sizes \citep{marshall1948raindrops}.

Conventional QPE pipelines require scan-by-scan decoding, calibration, resampling, and accumulation—all steps which must be repeated for each archive. Radar DataTree enables direct access to reflectivity fields aligned in time and space, significantly reducing overhead.

In a three-week, multi-radar QPE case study, we observed 70–150× performance gains in total compute time, with reproducible accumulation fields generated using Icechunk-managed updates and versioned datasets.

\begin{figure}[h]
  \centering
  \includegraphics[width=0.95\textwidth]{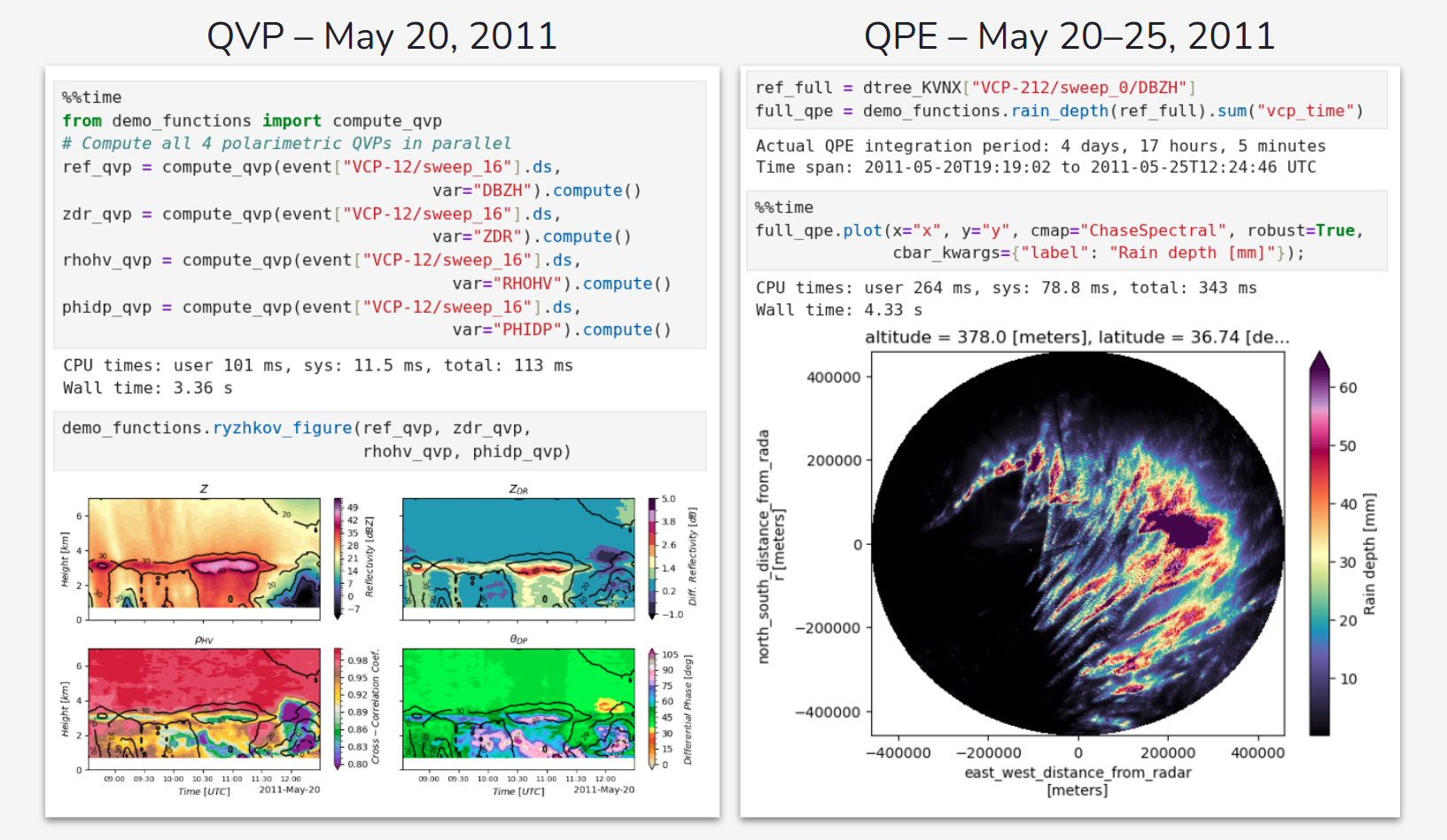}
  \caption{
    Interactive computation of two standard radar science products from the KVNX Radar DataTree using \texttt{xarray} and Dask. (\textbf{Left}) Quasi-Vertical Profiles (QVPs) of four polarimetric variables for VCP-12 on May 20, 2011, following the method of \citet{ryzhkov2016qvp}. Total compute time: 3.36~s. (\textbf{Right}) Quantitative Precipitation Estimation (QPE) by time-integrating radar reflectivity (DBZH) from VCP-212 sweeps over 4.7~days using a reflectivity–rainrate relation \citep{marshall1948raindrops}. Compute time: 4.33~s on a 10-worker Dask cluster.
  }
  \label{fig:qvp-qpe}
\end{figure}

Figure~\ref{fig:qvp-qpe} shows interactive computation of Quasi-Vertical Profiles (QVPs) and Quantitative Precipitation Estimation (QPE) directly from the Radar DataTree, computed with cloud-native tools. Both examples are available in the Radar DataTree Demo repository\footnote{\url{https://github.com/earth-mover/radar-data-demo}}, which demonstrates reproducible workflows using Earthmover infrastructure.

All parallel processing benchmarks were run on a 10-worker Dask cluster\footnote{
Cluster deployed using \texttt{coiled} with 10 \texttt{m6i.large} workers (2 vCPUs, 8~GiB RAM each) and a \texttt{m6i.xlarge} scheduler in \texttt{us-east-1}. Software environment: \texttt{nexrad-env}. Full dashboard: \url{https://cluster-gxyov.dask.host}.
}.

\subsection{Transactional Updates and Reproducibility}

Radar DataTree’s integration with Icechunk allows radar archives to support safe concurrent access and real-time ingestion. New volume scans can be appended as ACID-compliant transactions, enabling downstream analytics to operate on live data without reprocessing the entire archive.

In a validation experiment, a month-long dataset was constructed incrementally using daily data streams. Versioned Icechunk commits allowed for rollback and re-execution of QVP and QPE analyses with bitwise-identical results, confirming both reproducibility and provenance tracking.

\section{Conclusion and Impact}

Despite its scientific value, weather radar data has remained structurally underserved in modern data systems. While archives such as NEXRAD Level II, IDEAM Colombia, and FMI Finland are now publicly hosted in the cloud, they remain locked behind file-based workflows, legacy formats, and metadata inconsistencies.

Radar DataTree addresses this gap by introducing a dataset-level abstraction aligned with FAIR data principles and optimized for cloud-native environments. By extending FM-301 from individual scans to time-indexed collections, and combining \texttt{xarray.DataTree}, Zarr, and Icechunk, we offer a complete framework for structured, scalable radar data workflows.

\textbf{Intended audience.} This framework is designed for radar data scientists, meteorologists, atmospheric researchers, open science advocates, and data infrastructure engineers working at the intersection of environmental data and scalable computing.

% End of main content

\bibliographystyle{unsrtnat}
\bibliography{references}

\end{document}